\documentclass[a4paper]{article}

\usepackage{icrc2013}

\title{Study of the muon content of very high-energy EAS measured with  
the KASCADE-Grande observatory}

\shorttitle{Study of the muon contect of VHE EAS measured with KASCADE-Grande}

\authors{
J.C.~Arteaga-Vel\'azquez$^{2}$,
W.D.~Apel$^{1}$,
K.~Bekk$^{1}$,
M.~Bertaina$^{3}$,
J.~Bl\"umer$^{1,4}$,
H.~Bozdog$^{1}$,
I.M.~Brancus$^{5}$,
E.~Cantoni$^{3,6,a}$,
A.~Chiavassa$^{3}$,
F.~Cossavella$^{4,b}$,
C.~Curcio$^{3}$,
K.~Daumiller$^{1}$,
V.~de Souza$^{7}$,
F.~Di~Pierro$^{3}$,
P.~Doll$^{1}$,
R.~Engel$^{1}$,
J.~Engler$^{1}$,
B.~Fuchs$^{4}$,
D.~Fuhrmann$^{8,c}$,
H.J.~Gils$^{1}$,
R.~Glasstetter$^{8}$,
C.~Grupen$^{9}$,
A.~Haungs$^{1}$,
D.~Heck$^{1}$,
J.R.~H\"orandel$^{10}$,
D.~Huber$^{4}$,
T.~Huege$^{1}$,
K.-H.~Kampert$^{8}$,
D.~Kang$^{4}$, 
H.O.~Klages$^{1}$,
K.~Link$^{4}$, 
P.~{\L}uczak$^{11}$,
M.~Ludwig$^{4}$,
H.J.~Mathes$^{1}$,
H.J.~Mayer$^{1}$,
M.~Melissas$^{4}$,
J.~Milke$^{1}$,
B.~Mitrica$^{5}$,
C.~Morello$^{6}$,
J.~Oehlschl\"ager$^{1}$,
S.~Ostapchenko$^{1,d}$,
N.~Palmieri$^{4}$,
M.~Petcu$^{5}$,
T.~Pierog$^{1}$,
H.~Rebel$^{1}$,
M.~Roth$^{1}$,
H.~Schieler$^{1}$,
S.~Schoo$^{1}$,
F.G.~Schr\"oder$^{1}$,
O.~Sima$^{12}$,
G.~Toma$^{5}$,
G.C.~Trinchero$^{6}$,
H.~Ulrich$^{1}$,
A.~Weindl$^{1}$,
J.~Wochele$^{1}$,
J.~Zabierowski$^{11}$ \\
KASCADE-Grande Collaboration
}

\afiliations{
$^1$ Institut f\"ur Kernphysik, KIT - Karlsruher Institut f\"ur Technologie, Germany\\
$^2$ Universidad Michoacana, Instituto de F\'{\i}sica y Matem\'aticas, Morelia, Mexico\\
$^3$ Dipartimento di Fisica, Universit\`a degli Studi di Torino, Italy\\
$^4$ Institut f\"ur Experimentelle Kernphysik, KIT - Karlsruher Institut f\"ur Technologie, 
Germany\\
$^5$ National Institute of Physics and Nuclear Engineering, Bucharest, Romania\\
$^6$ Osservatorio Astrofisico di Torino, INAF Torino, Italy\\
$^7$ Universidade S$\tilde{a}$o Paulo, Instituto de F\'{\i}sica de S\~ao Carlos, Brasil\\
$^8$ Fachbereich Physik, Universit\"at Wuppertal, Germany\\
$^9$ Department of Physics, Siegen University, Germany\\
$^{10}$ Dept. of Astrophysics, Radboud University Nijmegen, The Netherlands\\
$^{11}$ National Centre for Nuclear Research, Department of Cosmic Ray Physics, Lodz, Poland\\
$^{12}$ Department of Physics, University of Bucharest, Bucharest, Romania\\
\scriptsize{
$^{a}$ now at: Istituto Nazionale di Ricerca Metrologia, INRIM, Torino; \\
$^{b}$ now at: Max-Planck-Institut Physik, M\"unchen, Germany; \\
$^{c}$ now at: University of Duisburg-Essen, Duisburg, Germany;  \\
$^{d}$ now at: University of Trondheim, Norway.
}
}

\email{arteaga@ifm.umich.mx}

\abstract{The KASCADE-Grande detector is an air-shower array devoted to the study of 
 primary cosmic rays with very high-energies ($E = 10^{16} - 10^{18} \, \mbox{eV}$). 
 The instrument is composed of different particle detector systems suitable for 
 the detailed study of the properties of Extensive Air Showers (EAS) developed by 
 cosmic rays in the atmosphere. Among the EAS observables studied with the detector, 
 the charged number of particles, the muon content (at different energy thresholds), 
 and the number of electrons are found. By comparing the measurements of these 
 air-shower parameters with the expectations from MC simulations, different hadronic 
 interaction models can be tested at the high-energy regime with the KASCADE-Grande 
 experiment. In this work, the results of a study on the evolution of the muon content 
 of EAS with zenith angle, performed with the KASCADE-Grande instrument, is presented. 
 Measurements are compared with predictions from MC simulations based on the QGSJET II, 
 QGSJET II-04, SIBYLL 2.1 and EPOS 1.99 hadronic interaction models. A mismatch between 
 experiment and simulations is observed. A similar problem is found for the evolution
 of the lateral distribution function of muons in the atmosphere.}

\keywords{Cosmic rays, hadron interaction models, KASCADE-Grande, muons}

\begin{document}
\maketitle

\section{Introduction}

 At very high-energies, cosmic rays are studied by observing the extensive
 air showers (EAS) that they produced in the atmosphere upon arrival to the 
 Earth. For this goal, EAS observatories are equipped with several kinds of 
 instruments that are used to study the various components of air showers 
 at different stages. The keys to the knowledge of the properties of 
 primary cosmic rays are to be found in the measured EAS observables. However,
 the interpretation of this information relies on the understanding of the
 physics behind the EAS. The production and development of air showers is 
 governed by particle physics, where an important source of uncertainty is 
 the role of hadronic interactions at the very high-energy regime. These 
 have a deep impact on the energy assigment and mass identification of 
 cosmic rays from the measured EAS data (see, for example, 
 \cite{K-unfold}).

 The lack of accelerator data at very high-energies and the difficulty of 
 solving the QCD equations in a non-perturbative region are the main
 obstacles to make accurate predictions at the high-energy regime when 
 hadronic interactions are involved. Nowadays, these difficulties are 
 handled by using phenomenological models calibrated at low energies 
 with data from man-made accelerators \cite{Knapp}.       

  At the highest energies, cosmic ray observatories can be used to  
 validate hadronic interaction models by comparing EAS predictions against 
 first-hand data from air-shower detectors, for example,
 about muons. Muons are produced in the decay of baryons and mesons generated 
 in hadronic collisions during the early stages of the EAS. They are very 
 penetrating particles and, as a consequence, they suffer less attenuation in 
 the atmosphere than the electromagnetic and hadronic components of the shower. 
 Therefore, muons can keep direct information from the properties of the 
 hadronic interactions at very high-energies. In this paper, a preliminary 
 study is performed to test the predictions of the QQGSJET II \cite{qgs}, 
 QGSJET II-04 \cite{qgs04}, SIBYLL 2.1 \cite{sibyll} and EPOS 1.99 \cite{epos} hadronic 
 interaction models about the muon content of EAS at very 
 high-energies. The research is based on the air-shower data collected with the 
 KASCADE-Grande experiment \cite{kg-NIM10} in the energy interval from $10^{16}$ to 
 $10^{18} \, \mbox{eV}$.

 \section{The KASCADE-Grande observatory}

 KASCADE-Grande is an air-shower multi-detector observatory dedicated to study 
 cosmic rays with energies $E = 10^{16} - 10^{18} \, \mbox{eV}$ \cite{kg-NIM10}. 
 The main part of the 
 experiment is a $0.5 \, \mbox{km}^2$ array of $37 \times 10 \, \mbox{m}^2$ plastic 
 scintillator detectors, called Grande, which is employed to measure the arrival times 
 and the density of charged particles at the shower front. This information is later
 used to reconstruct the impact point of the EAS core at ground and both the arrival 
 direction and the charged number of particles ($N_{ch}$) of the air 
 shower. Another important component of the observatory is the KASCADE muon array, 
 composed by $192 \times 3.2 \, \mbox{m}^2$ shielded scintillator detectors. The array
 has an energy threshold for vertical muons of $230 \, \mbox{MeV}$ and measures the
 muon densities of the EAS at ground level. The total number of muons, $N_\mu$, is 
 obtained from these measurements. 

 \begin{figure}[!t]
  \centering
  \includegraphics[height=2.0in, width=3.2in]{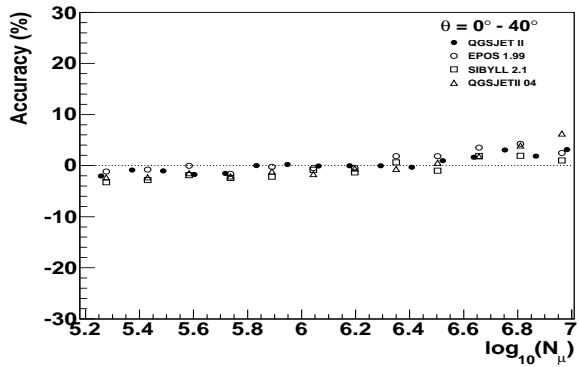}
  \caption{Muon systematics for the corrected EAS muon number assuming
  a mixed composition with $\gamma  = -3$. Results
  are presented for different hadronic interaction models.}
  \label{Fig01}
 \vspace{-1pc}
 \end{figure}

 \section{The data}

  All air shower simulations were performed with CORSIKA 
 \cite{cors} using Fluka \cite{fluka} to treat hadronic interactions in the low energy
 regime ($E < 200 \, \mbox{GeV}$). At high energies, the hadronic interaction models 
 QGSJET II, QGSJET II 04, SIBYLL 2.1 and EPOS v1.99, were employed. The response of the 
 detector was simulated with a GEANT 3.21 based code. Sets with energy spectra of the 
 form $E^{\gamma}$ were produced for a spectral index $\gamma = -2$ and afterwards 
 were weighted to have spectra with $\gamma = -2.8,-3,-3.2$. Sets for H, He, C, Si and 
 Fe cosmic-ray nuclei were simulated and were combined to reproduce a mixed composition 
 scenario (all single primaries in equal abundances). 
  
 Selection cuts were applied to both experimental and simulated data. They were
 chosen according to MC studies to avoid as much as possible the influence of
 systematic uncertainties on the reconstruction of EAS parameters. The
 selected data were composed of events with more than 11 triggered stations in
 Grande, shower cores inside a central area of $8 \times 10^{4} \, \mbox{m}^2$,
 core distance to the KASCADE center between $270$ and $440 \, \mbox{m}$, and arrival directions 
 confined to the zenith angle interval of $\Delta \theta = 0^\circ-40^\circ$. 
 These events were registered during  stable periods of data acquisition and passed
 successfully the standard reconstruction procedure of KASCADE-Grande \cite{kg-NIM10}. 
 Additionally, only showers with $\log N_\mu > 4.6$ were considered for this work.
 Both the experimental and simulated data were analyzed and reconstructed with 
 the same algorithms.  With the above quality cuts, the effective time of
 observation with KASCADE-Grande was equivalent to 1434 days. The threshold for 
 full efficiency depends slightly on the hadronic model, composition and arrival 
 direction, but in general is completely achieved above $\log N_\mu = 5.2$. 

 \begin{figure}[!t]
  \centering
  \includegraphics[height=2.0in, width=3.2in]{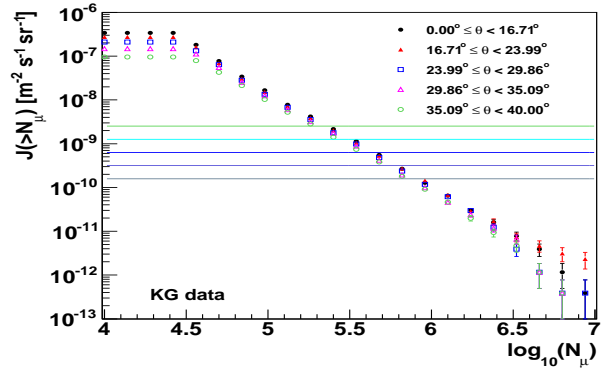}
  \caption{Integral muon fluxes for five zenith angle intervals derived 
  from the measurements with the KASCADE-Grande observatory. The muon
  correction function was already applied to the data. The CIC cuts employed
  in this work are shown as horizontal lines.}
  \label{Fig02}
 \vspace{-1pc}
 \end{figure}

 \section{Description of the analysis and results}
 
  Before starting the analysis, all experimental and simulated muon data was corrected 
 for systematic uncertainties in order to improve their accuracy. 
 That was done using a unique muon correction function derived from 
 MC simulations based on QGSJET II, assuming mixed composition and a spectral index 
 $\gamma = -3$. The employment of a single correction function is justified since 
 it was found that this function is nearly independent from the composition of cosmic 
 rays and from the hadronic interaction model. The function was parametrized with respect 
 to core position, azimuthal and zenithal angles, and muon size. After correction, mean 
 muon systematics become less than $6 \%$ for a mixed composition assumption ($6 \%$ and 
 $7\%$ for H and Fe, respectively) and have a mild dependence with the core position and 
 the muon size in the full efficiency regime (see, for example, Fig. \ref{Fig01}). This 
 is not a surprise, since selection cuts were carefully selected for this purpose.

 \begin{figure}[!t]
  \centering
  \includegraphics[height=2.0in, width=3.2in]{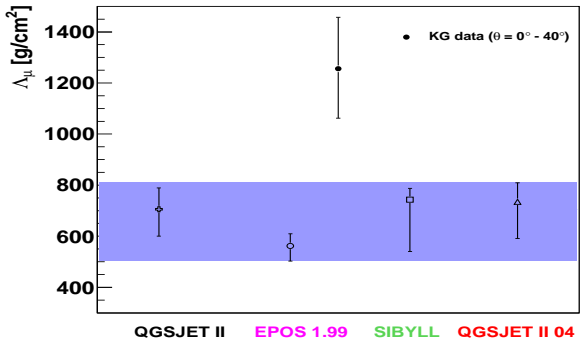}
  \caption{Muon attenuation lengths extracted from Monte Carlo (points inside shadowed 
  area) and experimental data (upper point). They are shown with the corresponding 
 systematic errors.}
  \label{Fig03}
 \vspace{-1pc}
 \end{figure}

  For a first test on the hadronic interaction models with the KASCADE-Grande muon data, 
 predictions on the evolution of the muon content with the arrival zenith angle of the 
 EAS were confronted with observations. The task was done comparing the expected and 
 observed values of the muon attenuation length, $\Lambda_\mu$. This quantity was 
 extracted by applying the Constant Intensity Cut (CIC) method to the integral
 muon spectra (see, e.g., Fig. \ref{Fig02}) as described in reference \cite{kg-ap12}. 
 Five CIC cuts were applied, here within the interval $\log_{10}[J_\mu/ \mbox{m}^2 \cdot 
 \mbox{s} \cdot \mbox{sr}] = [-9.8, -8.6]$, where maximum efficency and statistics 
 are achieved. From these cuts, muon attenuation curves, $\log_{10} N_\mu(\theta)$,
 are obtained and $\Lambda_\mu$ is extracted from a global fit to such a curves with 
 the formula
 \begin{equation}
   N_\mu = N^{0}_{\mu} \mbox{exp}[-X_0 \mbox{sec}(\theta)/\Lambda_\mu],
   \label{eqn1}
 \end{equation}
 where $X_0 = 1022 \, \mbox{g/cm}^2$ is the average atmospheric depth for
 vertical showers and $N^{0}_\mu$ is a normalization parameter to
 be determined for each attenuation curve.  The results for $\Lambda_\mu$ are 
 presented in Table \ref{table1} and Fig. \ref{Fig03} along with the
 corresponding systematics. The quoted values for $\Lambda_\mu$, in case of MC
 data, correspond to those data sets with a mixed composition assumption and 
 $\gamma = -3$. However, systematic uncertainties take into account the spreading 
 of the MC value of $\Lambda_\mu$ when the spectral index is modified 
 (considering $\gamma = -2.8$ and $-3.2$) and the cosmic ray 
 composition is changed. For the latter, the light and heavy composition hypotheses
 were considered. In case of the QGSJET II model, only protons were used for
 the light composition assumption, and iron nuclei for the heavy one. However,
 for the other hadronic interaction models, H and He nuclei, and Fe and Si,
 were employed with the abovementioned situations, respectively. Systematic errors
 for MC include also the variation found when the right correction
 function for the considered model is employed instead of that derived from
 QGSJET II. This excercise cannot be done with experimental data, since the
 real correction functions are unknown. Therefore, only the effect of 
 considering different correction functions derived from the hadronic interaction 
 models here employed was computed. 

 As observed from Table \ref{table1} and Fig. \ref{Fig03}, the experimental
 value of $\Lambda_\mu$ is well outside the area spanned by the MC values, 
 even considering the corresponding systematic errors. The smallest $\Delta 
 \Lambda_\mu$ difference between the experimental and MC values is found when 
 the observed value is compared with SIBYLL 2.1 ($\sim 2.6 \, \sigma_{\mbox{exp}}$)
 and the biggest one, when using EPOS 1.99 ($\sim 3.6 \, \sigma_{\mbox{exp}}$). 
 
 The differences between the measured and predicted $\Lambda_\mu$ are not the result
 of the application of the muon correction function on the data, since the
 discrepancies are still present even before using such a correction. Even more, they can 
 be tracked down to the differences in the evolution of the muon densities with the 
 arrival zenith angle, $theta$, as it will be shown below.

 \begin{figure}[!t]
  \centering
  \includegraphics[height=2.0in, width=3.2in]{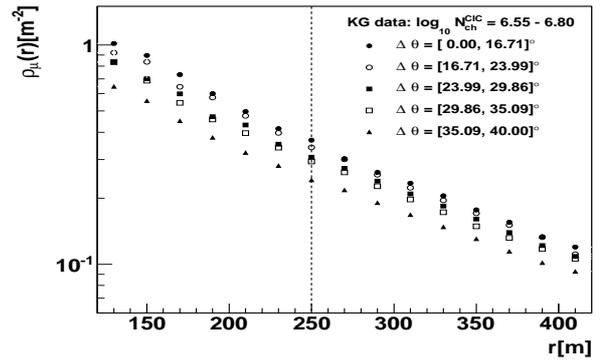}
  \caption{Measured lateral muon density distributions of EAS for
  data in the $N^{CIC}_{ch}$  interval from $7.04$ to  $7.28$. In the
  framework of QGSJET II, these distributions correspond to events with
  energy $\log_{10}(E/\mbox{eV}) = 16.9 - 17.1$, assuming a mixed composition 
  scenario. The vertical line is used to get the muon densities for different 
  $\theta$ intervals at a fixed radius.}
  \label{Fig04}
 \vspace{-1pc}
 \end{figure}

\begin{table*}[!t]
\footnotesize
\begin{center}
\begin{tabular}{l|c|c|c|c|c}
\hline 
  & QGSJETII & QGSJETII04 & EPOS 1.99 & SIBYLL 2.1 & KG data \\ 
\hline
 {\bf $\Lambda_\mu$ $(\mbox{g/cm}^2)$} &
 $705.95^{+82.93}_{-105.36}$ & $735.11^{+74.38}_{-143.75}$ & 
 $561.71^{+47.41}_{-59.20}$ & $743.16^{+44.35}_{-203.36}$& $1255.63^{+201.31}_{-193.59}$\\ 
 \hline
\textbf{Systematics ($\%$)}  &&&&&\\
Bin size  &$+6.60$&$+6.93$&$+3.95$&$-3.41$&$+6.79$\\
Global fit&$+4.37$&$+4.51$&$+4.59$&$+4.43$&$+5.60$\\
Narrower CIC interval &$-0.34$&$-1.90$&$-0.57$&$-0.98$&$-0.60$\\
Extended CIC interval &$+3.81$&$-1.01$&$+5.32$&$-7.87$&$-0.13$\\

Muon systematics      &$+0.04$&$-0.20$&$+0.21$&$-4.00$&$+1.97/-2.53$\\
Statistical fluctuations &$-0.07$&$+0.001$&$-0.04$&$+0.04$&$-1.37$\\
Core close to KASCADE &$-3.29$&$+2.40$&$+0.55$&$-4.33$&$-10.73$\\
Core far from KASCADE &$+0.29$&$-6.61$&$-3.12$&$-5.60$&$+11.89$\\
Broader zenith angle interval &$+1.98$&$+1.69$&$+0.90$&$-1.51$&$-2.42$\\
Seven CIC cuts &$-0.65$&$-0.45$&$-0.26$&$+0.45$&$+0.27$\\
Three CIC cuts &$+1.35$&$+0.52$&$+0.82$&$-2.50$&$+1.40$\\
Muon Correction function uncertainties &$+6.64$&$-0.40$&$+0.33$&$-2.88$&$-2.54$\\
$\gamma = -3.2$ &$+0.06$&$+5.00$&$+2.00$&$+3.97$&$-$\\
$\gamma = -2.8$ &$+1.35$&$-3.14$&$+0.35$&$-6.57$&$-$\\
Composition (Light)   &$+2.99$&$-13.19$&$-8.92$&$-22.44$&$-$\\
Composition (Heavy)   &$-13.86$&$-11.37$&$-0.52$&$-4.06$&$-$\\
Hadronic interaction model &$-$&$-0.40$&$+0.33$&$-2.88$&$+5.66/-8.34$\\

\textbf{Total}   &&&&&\\
                &$+11.75$&$+10.12$&$+8.44$&$+5.97$&$+16.03$\\
               &$-14.92$&$-19.55$&$-10.53$&$-27.36$&$-15.42$\\
\hline
\end{tabular}
\caption{Muon attenuation lengths extracted from Monte Carlo and experimental
      data. $\Lambda_\mu$ is presented along with its corresponding systematic 
      uncertainty. The respective contributions to the total systematic errors 
      are also shown.}
\label{table1}
\end{center}
 \vspace{-1.5pc}
\end{table*}

 The second test applied to the hadronic interaction models consisted of 
 confronting their predictions for the evolution of the muon density 
 distributions, $\rho_\mu(r)$, at the shower plane, with the angle $\theta$.  The
 study was focused on the energy range from $10^{16.2}$ to $10^{17} \, \mbox{eV}$,
 which roughly covers the $N_\mu$ interval from which $\Lambda_\mu$ was estimated. 
 Since data from different zenith angles is going to be compared, the CIC method 
 was applied again to proceed in a model independent way. However, in this case,
 it was used on the independent observable, $N_{ch}$, following \cite{kg-ap12}.  
 Then, $\Lambda_{ch}$ was estimated and employed to correct $N_{ch}$ for attenuation 
 effects in the atmosphere and to calculate the corresponding charged number of 
 particles, $N^{CIC}_{ch}$, that a given EAS would produced in case of being observed 
 at a reference angle, $\theta_{ref} = 22^{\circ}$. This angle corresponds to the mean 
 of the zenith angle distribution of the experimental data. Data was later classified 
 in different $N^{CIC}_{ch}$ intervals. For each of those intervals, the $\rho_\mu(r)$ 
 distributions of the five zenith angle intervals involved in this work were
 calculated. According to the CIC method, this data should show us how the lateral 
 muon density distributions of EAS of a given energy evolve with the atmospheric depth.
 In figure \ref{Fig04}, the $\rho_\mu(r)$ distributions for different $\theta$
 intervals and the range $N^{CIC}_{ch} = [7.04, 7.28]$ are presented. According
 to the QGSJET II model, these events correspond to air showers with energies in the interval
 from $\log_{10}(E/\mbox{eV}) = 16.9$ to $17.1$, assuming a mixed composition and 
 $\gamma = -3$.
  
\begin{figure}[!b]
  \centering
  \includegraphics[height=1.8in, width=3.2in]{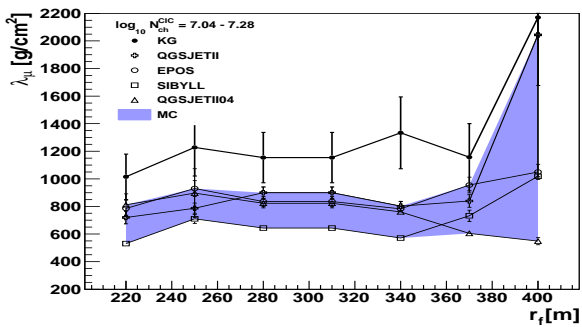}
  \caption{Measured muon absorption coefficients for both experimental and simulated
  data (assuming mixed composition) extracted for EAS in the interval  
  $N^{CIC}_{ch} = [7.04, 7.28]$.}
  \label{Fig05}
 \vspace{-1pc}
 \end{figure}

 Within each $N^{CIC}_{ch}$ interval, the $\rho_\mu(r)$ distributions are used to 
 obtain absorption curves for the muon density at different fixed distances, $r_f$, as 
 a function of $\sec (\theta)$. By fitting the above curves with the equation:
 \begin{equation}
   \rho_\mu (r_f) = \rho^{0}_{\mu} (r_f) \mbox{exp}[-X_0 \mbox{sec}(\theta)/\lambda_\mu (r_f)],
   \label{eqn2}
 \end{equation} 
 the absorption coefficient, $\lambda_\mu$, for the muon density at the shower front 
 is estimated as a function of $r_f$. Figure \ref{Fig05} shows the results of applying
 the above analysis to the experimental data of figure \ref{Fig04} along with the 
 expectations from MC models for the same $N^{CIC}_{ch}$ range under a mixed composition 
 assumption. Data are shown for $r_f = [220, 400] \, \mbox{m}$, where statistical 
 fluctuations of MC data are not so big. It is evident from figure \ref{Fig05} that the 
 evolution of experimental data in the atmosphere is  different from that predicted by the 
 hadronic interaction models studied in this work (at least for mixed compostion\footnote{
 In case of QGSJET II more statistics is available,
 so the excercise can be repeated for pure protons and iron nuclei. The result is that  
 experimental data are still outside the expected values for pure composition assumptions}). 
 Even more, by comparing these results with those from other $N^{CIC}_{ch}$  intervals, it 
 is found that the differences between measurements and predictions decrease with the 
 energy of the EAS. The latter effect seems
 to be more remarkable for particles close to the shower core, where a better agreement with
 simulations appears. These results have no influence from the muon correction functions, 
 since they were not applied on the muon densities. 

 Finally, the magnitude of the muon densities were compared with the predictions of 
 the different models. In general, for $\theta = [0^{\circ}, 40^{\circ}]$, previous
 observations were confirmed that the experimental $\rho_\mu(r)$ distributions 
 are well inside the limits predicted by the models for protons and iron 
 nuclei \cite{Souza}.

 \section{Conclusions}
 
 Preliminary analyses performed with the EAS data measured by KASCADE-Grande point out 
 that the observed evolution of the muon content of EAS with the atmospheric depth
 is not described by the hadronic interaction models QGSJET II, EPOS 1.99, QGSJET II-04
 and SIBYLL 2.1. These differences can be tracked down to discrepancies between
 the experimental and expected evolution of the muon density distributions of EAS in 
 the atmosphere. To corroborate or refute these findings, additional tests will be 
 performed on independent muon data collected by additional detector systems belonging
 to the experiment, for example, with the Muon Tracking Detector. 

\vspace*{0.5cm}
\footnotesize{{\bf Acknowledgment:}{KASCADE-Grande is supported by the BMBF of Germany, 
the MIUR and INAF of Italy, the Polish Ministry of Science and Higher Education (in part 
by grant for 2009-2011) and the Romanian Authority for Scientific Research. J.C.A.V. 
acknowledges the partial support from CONACYT and the Universidad Michoacana.}}


\begin{thebibliography}{}

\bibitem{K-unfold} W.-D. Apel et al., Astropar. Physics 24 (2005) 1-25
\bibitem{Knapp} J. Knapp et al., Astrop. Phys. 19, 77 (2003).
\bibitem{qgs} 
S.S.~Ostapchenko, Nucl. Phys. B (Proc. Suppl.) {\bf 151} (2006) 143\&147;
S.~Ostapchenko, Phys. Rev. D 74, 014026 (2006). 
\bibitem{qgs04}
S.S. Ostapchenko, Phys. Rev. D {\bf 83} (2011) 014018.
\bibitem{sibyll}
E.J. Ahn et al., Phys. Rev D 80, 094003 (2009).
\bibitem{epos}
T. Pierog et al., Report FZKA 7516, Forschungszentrum Karlsruhe 133, (2009).
\bibitem{kg-NIM10}
W.-D. Apel et al., NIM A {\bf 620}, 202 (2010).
\bibitem{cors}
D. Heck et al., Report FZKA 6019, Forschungszentrum Karlsruhe (1998).
\bibitem{fluka}   
A.~Fass\`o et al., Report CERN-2005-10, INFN/TC-05/11, SLAC-R-773 (2005).
\bibitem{kg-ap12}
W.-D. Apel et al., Astroparticle Phys. {\bf 36}, 183 (2012).
\bibitem{Souza}
V. Souza et al., Proc. of the 32nd ICRC, Beijing (2011), ID 0953.
\end{thebibliography}
\end{document}